\documentclass[12pt,a4paper]{article}
\usepackage{amsmath}
\usepackage{graphicx,amsmath,amssymb}

\usepackage{cite}
\usepackage{graphicx,amsmath,amssymb}

\DeclareMathAlphabet{\mathpzc}{OT1}{pzc}{m}{it}

\date{}
{ 
\title{ \bf  \Large  The hard to soft Pomeron transition in small $x$ DIS data using optimal renormalization} 

\author{  Martin Hentschinski$^{1}$, Agust{\' \i}n Sabio Vera$^{2,3}$, Clara Salas$^{2,3}$ 
\bigskip \\ { \normalsize
 $^1$ Brookhaven National Laboratory, Upton, NY 11973, USA}\\ \normalsize
$^2$ Instituto de F{\' \i}sica Te{\' o}rica UAM/CSIC 
and   Universidad Aut{\' o}noma de Madrid, \\  \normalsize   
E-28049 Madrid, Spain\\  \normalsize
$^3$ Physics Department, CERN, CH-1211 Gen\'eve 23, Switzerland
}

}
\begin{document} 

\maketitle

\begin{abstract}
  We show that it is possible to describe the effective Pomeron
  intercept, determined from the HERA Deep Inelastic Scattering data
  at small values of Bjorken $x$, using next-to-leading order BFKL
  evolution together with collinear improvements.  To obtain a good
  description over the whole range of $Q^2$ we use a non-Abelian
  physical renormalization scheme with BLM optimal scale, combined
  with a parametrization of the running coupling in the infrared
  region.
\end{abstract}

\section{Introduction \&  theoretical approach}

The description of Deep Inelastic Scattering (DIS) data for the
structure function $F_2$ in different regions of Bjorken $x$ and
virtuality of the photon $Q^2$ is one of the classical problems in
perturbative QCD. The literature on the subject is very large (see,
{\it e.g.}, the reviews in
Ref.~\cite{Dittmar:2005ed}
). In the present letter we are interested in regions with low values
of $x$ and revisit the theoretical approach to the problem using the
next-to-leading order (NLO)~\cite{Fadin:1998py} BFKL~\cite{BFKL1}
equation together with collinear improvements. We find that, in order
to get a good description over the full range of $Q^2$, we can use
optimal renormalization schemes. In this work we focus on indicating
which are the most important theoretical aspects which drive the bulk
of our results.  It is possible to introduce subleading refinements in
our calculation which make our predictions even closer to the data and
will be presented elsewhere.

Let us first review some well-known formulas for DIS in order to set
the ground for our treatment of the small $x$ resummation.  In DIS the
cross section is written in terms of the structure functions $F_2$ and
$F_L$ in the form
\begin{eqnarray}
\frac{d^2 \sigma}{ d x \, d Q^2} &=& \frac{2 \pi \alpha^2}{x Q^4} \left\lbrace [ 1+(1-y)^2 ]\, F_2(x,Q^2) - y^2 F_L(x,Q^2)\right\rbrace,
\end{eqnarray}
where $x$ and $y$ are the  dimensionless Bjorken variables, $Q^2$ the photon's virtuality and $\alpha$ the electromagnetic constant. More explicitly, for the structure functions, in terms of transverse and longitudinal polarizations of the photon, we have
\begin{eqnarray}
F_2 (x,Q^2) = \frac{Q^2}{4 \pi^2 \alpha} [\sigma_T (x,Q^2) +\sigma_L (x,Q^2)], 
\, \, F_L (x,Q^2) = \frac{Q^2}{4 \pi^2 \alpha} \sigma_L (x,Q^2),
\end{eqnarray}
where $\sigma_{L,T}$  is the cross-section for the scattering of a
transverse (longitudinal) polarized virtual photon on the proton.  At
large center-of-mass energy $\sqrt{s}$, which corresponds to the small $x \simeq
Q^2/s$ limit, high energy factorization makes it possible to write
the structure functions $F_I$, $I = 2, L$  in the form
\begin{eqnarray}
F_I  (x, Q^2)&=& \frac{1}{(2 \pi)^4} \int \frac{d^2 {\bf q}_\perp}{q^2} \int \frac{d^2 {\bf p}_\perp}{p^2} 
\Phi_I \left(q,Q^2\right) \Phi_P \left(p,Q^2_0\right) {\cal F} \left(s, q , p\right),
\end{eqnarray}
where all the integrations take place in the two-dimensional transverse momenta space with  $q = \sqrt{{\bf q}_\perp^2}$. The proton ($\Phi_P $) and photon ($\Phi_I$) impact factors are functions which are dominated by ${\cal O} (Q_0)$ and ${\cal O} (Q)$ transverse scales, respectively.  Note that the dependence of $\Phi_I$ on the photon virtuality can be calculated in perturbation theory. This is not the case for $\Phi_P $ whose dependence on the non-perturbative scale 
$Q_0 \simeq \Lambda_{\rm QCD}$ can only be modeled. 

If $Q^2$ was a scale similar to $Q_0^2$ then the gluon Green's function
${\cal F}$, which corresponds to the solution of the BFKL equation,
would be written as
\begin{eqnarray}
{\cal F} (s, q , p) &=& \frac{1}{2 \pi q \, p} \int \frac{d \omega}{2 \pi i} \int \frac{d \gamma}{2 \pi i} 
\left(\frac{q^2}{p^2}\right)^{\gamma-\frac{1}{2}} \left(\frac{s}{q \, p}\right)^\omega \frac{1}{\omega- \bar{\alpha}_s \chi_0\left(\gamma\right)},
\label{GGF1}
\end{eqnarray}
with $\bar{\alpha}_s = \alpha_s N_c / \pi$ and $\chi_0(\gamma) = 2 \psi(1) - \psi(\gamma)-\psi(1-\gamma)$  in a leading 
order (LO) approximation, which resums $\bar{\alpha}_s^n \log^n{s}$ terms to all-orders in the strong coupling.  $\psi(\gamma)$ is the logarithmic derivative of the Euler Gamma function. In DIS, however, $Q^2 \gg Q_0^2$ and this expression should be written in a form consistent with 
the resummation of $\bar{\alpha}_s \log{(1/x)}$ contributions:
\begin{eqnarray}
{\cal F} (s, q , p) &=& \frac{1}{2 \pi q^2} \int \frac{d \omega}{2 \pi i} \int \frac{d \gamma}{2 \pi i} 
 \left(\frac{{q^2}}{p^2}\right)^{\gamma} \left(\frac{s}{q^2}\right)^\omega \frac{1}{\omega- \bar{\alpha}_s \chi_0\left(\gamma - \frac{\omega}{2}\right)}.
\end{eqnarray}
It is well-known that the zeros of the denominator in the integrand generate in the limits $\gamma \to 0,1$ all-orders terms not compatible with DGLAP evolution~\cite{Salam:1998tj,Vera:2005jt}. The first of these pieces (${\cal O} (\alpha_s^2)$) is removed when the NLO correction to the BFKL kernel is taken into account but not the 
higher order ones, which remain and are numerically important. A scheme to eliminate these spurious contributions~\cite{Salam:1998tj}, in a 
nutshell, consists of using a modified BFKL kernel in Eq.~(\ref{GGF1}) where we essentially introduce the change $\chi_0(\gamma) \to \chi_0(\gamma+\omega/2)$.   

Let us present now in a precise manner our procedure to include the NLO corrections and collinear improvements. The NLO expansion of the BFKL kernel in terms of poles at $\gamma =0,1$ reads
\begin{eqnarray}
\omega &=& \bar{\alpha}_s \chi_0 (\gamma-\frac{\omega}{2}) +\bar{\alpha}_s^2 \chi_1 (\gamma) \nonumber\\
&=&  \bar{\alpha}_s \chi_0 (\gamma) +\bar{\alpha}_s^2 \chi_1 (\gamma) 
- \frac{1}{2} \bar{\alpha}_s^2 {\chi_0}'  (\gamma) \chi_0 (\gamma) + {\cal O} (\bar{\alpha}_s^3) \nonumber\\
&\simeq& \frac{\bar{\alpha}_s}{\gamma} + \bar{\alpha}_s^2 \left(\frac{a}{\gamma} + \frac{b}{\gamma^2} - \frac{1}{2 \gamma^3}\right)
+ \frac{\bar{\alpha}_s}{1-\gamma} +  \frac{\bar{\alpha}_s^2}{2\gamma^3} -  \frac{\bar{\alpha}_s^2}{2 (1-\gamma)^3}
\nonumber\\
&+& \bar{\alpha}_s^2 \left[\frac{a}{1-\gamma} + \frac{b}{(1-\gamma)^2} - \frac{1}{2 (1-\gamma)^3}\right]
+ {\cal O} (\bar{\alpha}_s^3), 
\label{kerneldis}
\end{eqnarray}
where ${\chi_0}'  (\gamma) = \psi' (1-\gamma) - \psi' (\gamma)$. Now, as we have explained before, we resum  in the Regge region  ($Q^2 \simeq Q_0^2$) collinear logarithms by introducing a shift of the general form~\cite{Salam:1998tj,Vera:2005jt}
\begin{eqnarray}
\omega = \bar{\alpha}_s (1+A \bar{\alpha}_s) \left[2 \psi (1) - \psi \left(\gamma + \frac{\omega}{2} + B \bar{\alpha}_s\right)- \psi \left(1-\gamma + \frac{\omega}{2} + B \bar{\alpha}_s\right)\right].
\end{eqnarray}
 When translated into the DIS limit ($Q^2 \gg Q_0^2$) this expression is to be replaced by
\begin{eqnarray}
\omega &=& \bar{\alpha}_s (1+A \bar{\alpha}_s) \left[2 \psi (1) - \psi \left(\gamma + B \bar{\alpha}_s\right) 
- \psi \left(1-\gamma + \omega + B \bar{\alpha}_s\right)\right] \nonumber\\
&&\hspace{-1.7cm}= \bar{\alpha}_s (1+A \bar{\alpha}_s) \sum_{m=0}^\infty \left(\frac{1}{\gamma+m+B \bar{\alpha}_s}
+\frac{1}{1-\gamma+m+\omega+ B \bar{\alpha}_s} - \frac{2}{m+1}\right).
\end{eqnarray}
It is possible to
get a very good approximation to the solution of this equation
(certainly within the uncertainty of the resummation scheme) by
breaking its transcendentality and solving it pole by pole and summing
up the different solutions.  This procedure was proposed in  Ref.~\cite{Vera:2005jt}. In terms of (anti-)collinear poles we
obtain
\begin{eqnarray}
\omega &=& \sum_{m=0}^\infty \Bigg\{ \bar{\alpha}_s (1+A \bar{\alpha}_s) \left(\frac{1}{\gamma+m+B \bar{\alpha}_s}- \frac{2}{m+1}\right) \nonumber\\
&+& \frac{1}{2} \left(\gamma - 1 -m - B \bar{\alpha}_s + \sqrt{(\gamma-1-m-B \bar{\alpha}_s)^2 + 4 
\bar{\alpha}_s (1+ A \bar{\alpha}_s)}\right)\Bigg\} \nonumber\\
&=& \sum_{m=0}^\infty \Bigg\{\bar{\alpha}_s 
\left(\frac{1}{\gamma+m}+\frac{1}{1-\gamma+m}- \frac{2}{m+1}\right) \nonumber\\
&+& \bar{\alpha}_s^2 \Bigg(\frac{A}{\gamma+m}+\frac{A}{1-\gamma+m}-\frac{B}{(\gamma+m)^2}-\frac{B}{(1-\gamma+m)^2}
\nonumber\\
&-& \frac{1}{(1+m-\gamma)^3}- \frac{2 A}{m+1}\Bigg)\Bigg\} + {\cal O} (\bar{\alpha}_s^3).
\end{eqnarray}
In order to match the NLO poles in Eq.~(\ref{kerneldis}) we need to fix $A=a$ and $B=-b$. Keeping the LO and NLO kernels unmodified and introducing only higher orders corrections, our collinearly improved BFKL kernel then simply reads
\begin{eqnarray}
{\chi}(\gamma) &=& \bar{\alpha}_s \chi_0 (\gamma) +\bar{\alpha}_s^2 \chi_1 (\gamma) 
- \frac{1}{2} \bar{\alpha}_s^2 {\chi_0}'  (\gamma) \chi_0 (\gamma) 
+ \chi_{\text{RG}}(\bar{\alpha}_s, \gamma, a, b) ,
\end{eqnarray}
with
\begin{eqnarray}
\label{eq:RG}
&& \chi_{\text{RG}}(\bar{\alpha}_s, \gamma, a, b)
\,\, =  \,\,\bar{\alpha}_s (1+ a \bar{\alpha}_s) \left(\psi(\gamma) 
- \psi (\gamma-b \bar{\alpha}_s)\right) \nonumber\\
&& \qquad \quad - \frac{\bar{\alpha}_s^2}{2} \psi'' (1-\gamma)  - b \bar{\alpha}_s^2 \frac{\pi^2}{\sin^2{(\pi \gamma)}}
+ \frac{1}{2} \sum_{m=0}^\infty \Bigg(\gamma-1-m+b \bar{\alpha}_s  \nonumber\\
&&
\qquad \quad - \frac{2 \bar{\alpha}_s (1+a \bar{\alpha}_s)}{1-\gamma+m}
+ \sqrt{(\gamma-1-m+b \bar{\alpha}_s)^2+ 4 \bar{\alpha}_s (1+a \bar{\alpha}_s)} \Bigg).
\end{eqnarray}
For the NLO  kernel,
\begin{eqnarray}
\label{chi1NLO}
\chi_1 (\gamma) & =& {\cal S} \chi_0 (\gamma) - \frac{\beta_0}{8 N_c}   \chi_0^2(\gamma)
+ \frac{ \Psi ''(\gamma) + \Psi''(1-\gamma)- \phi(\gamma)-\phi (1-\gamma) }{4} \nonumber  \\
&& \hspace{-1.8cm} - \frac{\pi^2 \cos{(\pi \gamma)}}{4 \sin^2{(\pi \gamma)}(1-2\gamma)}
\left[3+\left(1+\frac{n_f}{N_c^3}\right) \frac{2+3\gamma(1-\gamma)}{(3-2\gamma)(1+2\gamma)}\right] 
+ \frac{3}{2} \zeta(3) ,
\end{eqnarray}
with ${\cal S} = (4-\pi^2 + 5 \beta_0/N_c)/12$,  $\beta_0 =  \left(\frac{11}{3}N_c - \frac{2 }{3} n_f \right)$ and
\begin{eqnarray}
\phi(\gamma) + \phi (1-\gamma) &=& \nonumber\\
&& \hspace{-3cm}\sum_{m=0}^\infty \left(\frac{1}{\gamma+m}+\frac{1}{1-\gamma+m}\right)
\left[\Psi'\left(1+\frac{m}{2}\right)-\Psi'\left(\frac{1+m}{2}\right)\right],
\end{eqnarray}
we obtain for  the coefficients
\begin{eqnarray}
a &=& \frac{5}{12} \frac{\beta_0}{N_c} - \frac{13}{36} \frac{n_f}{N_c^3}- \frac{55}{36}, \hspace{1cm} 
b ~=~ - \frac{1}{8} \frac{\beta_0}{N_c} - \frac{n_f}{6N_c^3}- \frac{11}{12}.
\end{eqnarray}
As we have already indicated, the non-perturbative proton impact factor has to be modeled. We take the functional form 
\begin{eqnarray}
\label{eq:protoN}
\Phi_P \left(p,Q_0^2\right) &=& { {\cal C}} \left(\frac{p^2}{Q_0^2}\right)^\delta e^{-\frac{p^2}{Q_0^2}},
\end{eqnarray}
which introduces three independent free parameters and has a maximum at $p^2 = \delta \, Q_0^2$. Its  representation in $\gamma$ space reads
\begin{eqnarray}
\int \frac{d^2 p}{p^2}  \Phi_P \left(p,Q_0^2\right) (p^2)^{-\gamma} &=& \pi \, {\cal C} \, \Gamma(\delta-\gamma) (Q_0^2)^{-\gamma}.
\end{eqnarray}
In the present work we choose to keep the treatment of the impact factors as simple as possible in order to focus on the 
behaviour of the gluon Green's function. Having this philosophy in mind, we work with the LO photon impact factor which can be 
written in the form (directly in $\nu = i(1/2 - \gamma)$ space)
\begin{eqnarray}
\int \frac{d^2 q}{q^2}  \Phi_I \left(q,Q^2\right) \left({q^2 \over Q^2}\right)^{\gamma-1} =\;\; \alpha\, \bar{\alpha}_s \,\pi^4 
\sum_{q=1}^{n_f} e_q^2 \, \frac{\Omega_I(\nu)}{\nu+ \nu^3}\, {\rm sech}{(\pi \nu)}\, \tanh{(\pi \nu)}\; ,
\end{eqnarray}
where $\Omega_2 = (11+12 \nu^2)/8$ and $\Omega_L =  \nu^2+ 1/4$.

So far we have not included in the NLO kernel those terms breaking scale invariance, directly linked to the 
running of the coupling. They appear as a differential operator in $\nu$ space which acts on the impact factors (for a similar analysis see Ref.~\cite{Vera:2007dr}). 
Exponentiating only the scale invariant LO and NLO terms in the kernel, the structure functions can be written as
\begin{eqnarray}
F_I (x,Q^2) &=& {\cal D} \int_{-\infty}^\infty d \nu \,  {x}^{-{\chi} \left(\frac{1}{2} + i \nu \right)} c_I (\nu) c_P (\nu) 
\Bigg\{1\nonumber\\
&&\hspace{-1.6cm}+ \bar{\alpha}_s^2 \log{\left(\frac{1}{x}\right)} \frac{\beta_0}{8 N_c} \chi_0 \left(\frac{1}{2} + i \nu \right) \left[\log{(\mu^4)}+i \frac{d}{d \nu}
\log{\left(\frac{c_I(\nu)}{c_P (\nu)}\right)}\right]\Bigg\},\quad
\end{eqnarray}
where we have gathered  different constants in ${\cal D}$ and $\mu$ denotes the renormalization scale at which the QCD coupling is evaluated. 
 Since 
\begin{eqnarray}
c_I(\nu) &=& (Q^2)^{\frac{1}{2}+i \nu}\;  \frac{\Omega_I(\nu)}{\nu+ \nu^3}\; {\rm sech}{(\pi \nu)} \tanh{(\pi \nu)},\\
c_P (\nu) &=& \Gamma \left(\delta-\frac{1}{2}-i \nu\right) (Q_0^2)^{-\frac{1}{2}-i \nu},
\end{eqnarray}
we can write our final expression in the form
\begin{eqnarray}
F_I (x,Q^2) &=& {\cal D} \int_{-\infty}^\infty d \nu \,  x^{-{\chi} \left(\frac{1}{2} + i\nu \right)} c_I (\nu) c_P (\nu) 
\Bigg\{1\nonumber\\
&& \hspace{-1.5cm}+ \bar{\alpha}_s^2 \log{\left(\frac{1}{x}\right)} \frac{\beta_0}{8 N_c} \chi_0 \left(\frac{1}{2} + i \nu\right) 
\Bigg[-\log{\left(\frac{Q^2 Q_0^2}{\mu^4}\right)}- \psi \left(\delta-\frac{1}{2}-i \nu\right) \nonumber\\
 && \hspace{-1.5cm}+  i \Bigg(\pi {\rm coth} (\pi \nu)-2 \pi \tanh{(\pi \nu)}- M_I (\nu)\Bigg)\Bigg]\Bigg\},
\label{Frho}
\end{eqnarray}
where
\begin{eqnarray}
M_2 (\nu) &=& \frac{11+21 \nu^2+12 \nu^4}{\nu (1+\nu^2)(11+12 \nu^2)}, 
\hspace{.5cm}
M_L (\nu) ~=~ \frac{1 - \nu^2+ 4 \nu^4}{\nu(1+5 \nu^2+4 \nu^4)}.
\end{eqnarray}
Although we have included all the ingredients needed to calculate $F_L$, 
we leave a comparison to experimental data for this observable to  
future work and focus in the following on $F_2$.

\section{\hspace{-.25cm}Running coupling \& optimal renormalization}
\label{RunningScheme}

Although there is some freedom in the treatment of the running of the coupling, 
it is natural to  remove the $\mu$ dependent logarithm in the second line of
Eq.~(\ref{Frho}) making the replacement 
\begin{eqnarray}
\bar{\alpha}_s - \bar{\alpha}_s^2 \frac{\beta_0}{8 N_c} \log{\left(\frac{Q^2 Q_0^2}{\mu^4}\right)} \,\,\, \longrightarrow \,\,\, 
\bar{\alpha}_s \left(Q Q_0\right),
\label{RC1}
\end{eqnarray}
and use this resummed coupling throughout our calculations. 
We are interested in the comparison of our approach with DIS data in
the small $x$ region. In this letter we focus on the  description of the $Q^2$ dependence of the
well-known effective  intercept $\lambda (Q^2)$, which can
be obtained from experimental DIS  data in the region $x < 10^{-2}$ through
a  parametrization of the structure function of the form $F_2 (x,Q^2) = c(Q^2) x^{- \lambda (Q^2)}$.   The intercept $\lambda (Q^2)$ is ${\cal O}(0.3)$ at large values of $Q^2$ and ${\cal O}(0.1)$ at low
values, closer to the confinement region. This can be qualitatively
interpreted as a smooth transition from hard to soft Pomeron
exchange. When trying to  describe  these data with our approach we have
found that it is rather difficult to get  good agreement over the full
range of $1 \, {\rm GeV}^2 < Q^2 < 300 \, {\rm GeV}^2 $. Somehow it is
needed to introduce some new ideas related to the infrared region. A
recent very interesting possibility is that proposed by Kowalski,
Lipatov, Ross and Watt in Ref.~\cite{Kowalski:2010ue}.  Alternatively, we
have found that moving from the $\overline{\rm MS}$ scheme to
renormalization schemes inspired by the existence of a possible
infrared fixed point significantly helps in generating a natural fit for
$\lambda (Q^2)$, in the sense of having sensible values for the two free
parameters in our calculation which affect this observable: $\delta$
and $Q_0$ in the proton impact factor. Here we are guided by having a
proton impact factor which should be dominated by the infrared
region. In the following we provide some details of our findings.

The first evaluation of the BFKL Pomeron intercept in non-Abelian
physical renormalization schemes using the Brodsky-Lepage-Mackenzie
(BLM) optimal scale setting~\cite{Brodsky:1982gc} was performed in
Ref.~\cite{Brodsky:1998kn} in the context of virtual photon-photon
scattering. We will use the same procedure in our calculation. The
pieces of the BFKL kernel at NLO proportional to $\beta_0$ are
isolated and absorbed in a new definition of the running coupling in
such a way that all vacuum polarization effects from the $\beta_0$
function are resummed, {\it i.e.},
\begin{eqnarray}
\label{eq:BLM}
\tilde{\alpha}_s \left(Q Q_0, \gamma \right) &=& \frac{4 N_c}{\beta_0 \left[\log{\left(\frac{Q Q_0}{ \Lambda^2}\right)} 
+\frac{1}{2} \chi_0 (\gamma) - \frac{5}{3} +2 \left(1+ \frac{2}{3} Y\right)\right]},
\end{eqnarray}
where we are using the momentum space (MOM) physical renormalization scheme based on a symmetric triple gluon vertex~\cite{Celmaster:1979km} with 
$Y \simeq 2.343907$ and gauge parameter $\xi =3$ (our results are very weakly dependent on this choice). This scheme is more suited to the BFKL context since there are large non-Abelian contributions to the kernel. The 
replacements we need in our kernel in order to introduce this new scheme are $\bar{\alpha}_s \left(Q Q_0\right) \to \tilde{\alpha}_s \left(Q Q_0\right)$ in Eq.(\ref{RC1}) and $\chi_1 (\gamma) \to \tilde{\chi}_1 (\gamma)$ in Eq.~(\ref{chi1NLO}) together with the corresponding adjustments for the coefficients $a, b \to \tilde{a}, \tilde{b}$ which enter Eq.~(\ref{eq:RG}). They read
\begin{eqnarray}
\tilde{\chi}_1 (\gamma) &=& \tilde{\cal S} \chi_0 (\gamma) + \frac{3}{2} \zeta(3)
+ \frac{ \Psi ''(\gamma) + \Psi''(1-\gamma)- \phi(\gamma)-\phi (1-\gamma) }{4} \nonumber \\
&-& \frac{\pi^2 \cos{(\pi \gamma)}}{4 \sin^2{(\pi \gamma)}(1-2\gamma)}
\left[3+\left(1+\frac{n_f}{N_c^3}\right) \frac{2+3\gamma(1-\gamma)}{(3-2\gamma)(1+2\gamma)}\right] \nonumber\\
&+&\frac{1}{8} \left[\frac{3}{2} (Y-1) \xi
   +\left(1-\frac{Y}{3}\right) \xi ^2+\frac{17 Y}{2}-\frac{\xi ^3}{6}\right] \chi_0 (\gamma), \\
\tilde{a} &=&  - \frac{13}{36} \frac{n_f}{N_c^3}- \frac{55}{36} + \frac{3 Y - 3}{16}\xi + \frac{3 - Y}{24} \xi^2 - \frac{1}{48}\xi^3 + \frac{17}{16}Y
 \\
\tilde{b} &=& - \frac{n_f}{6N_c^3}- \frac{11}{12},
\end{eqnarray}
where $\tilde{\cal S} = (4-\pi^2)/12$. 

\begin{figure}[htbp]
 \parbox{.5\textwidth}{ \includegraphics[width=.5\textwidth,angle=0]{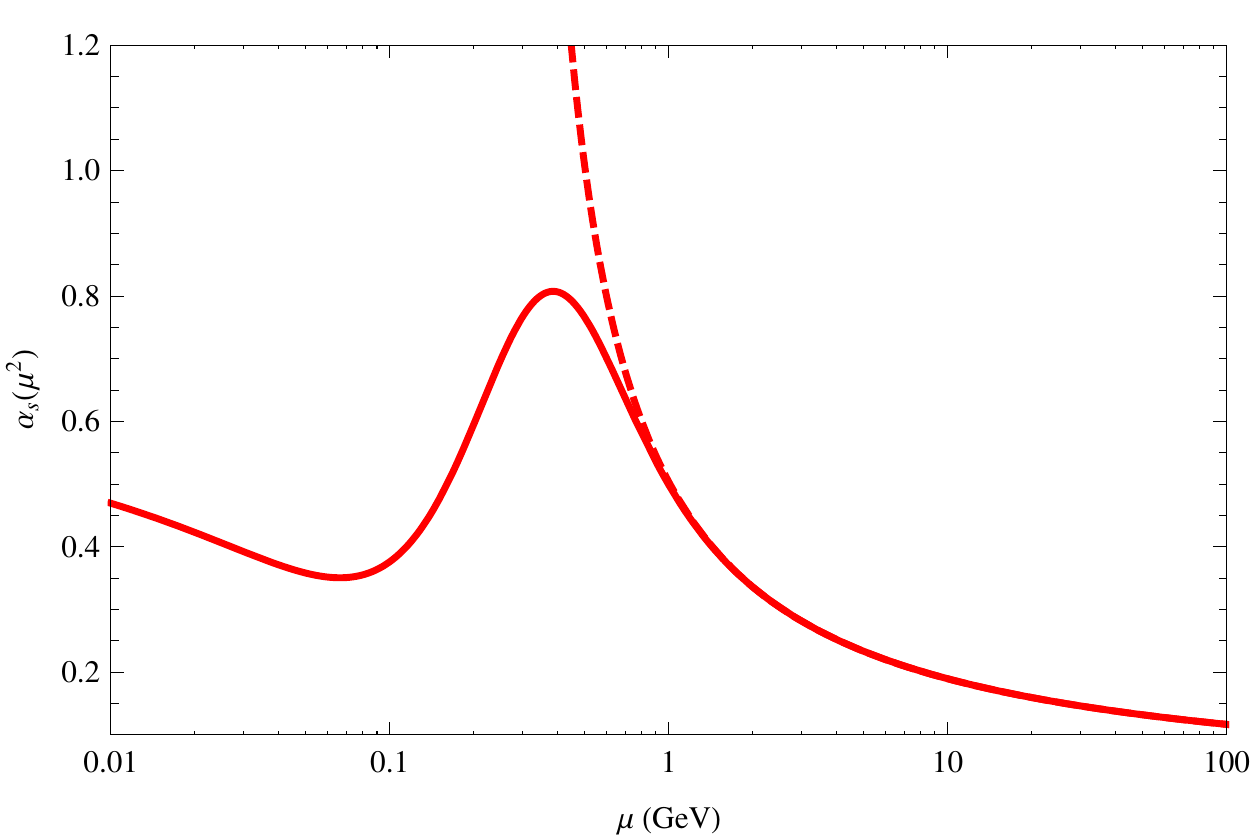}}
 \parbox{.5\textwidth}{\includegraphics[width=.5\textwidth,angle=0]{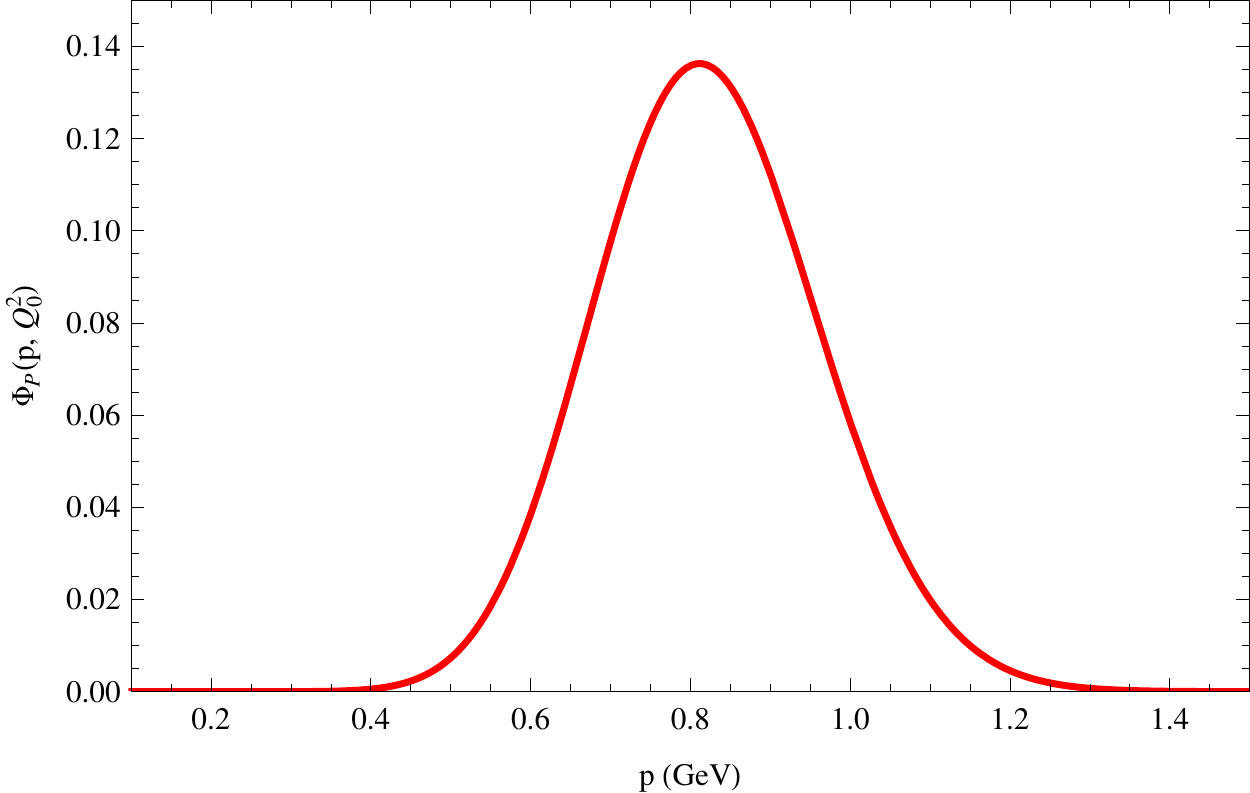}}
  \caption{{\small Left: model for the running coupling with freezing in the infrared (solid line) and leading order running coupling with Landau pole (dashed line) for $n_f = 3$ and $\Lambda=0.25$ GeV. Right: proton impact factor in momentum space with $\mathcal{C} = 1/{\Gamma(1 + \delta)}$ and $\delta, Q_0$ with the values used for the comparison to DIS data.}}
   \label{BothRCPIF}
\end{figure}

In order to access regions with $Q^2 \simeq 1 \,  {\rm GeV}^2$, we use a simple parametrization of the running coupling introduced by Webber in Ref.~\cite{Webber:1998um}: 
 \begin{eqnarray}
\alpha_s \left(\mu^2\right) =  \frac{4\pi}{\beta_0\ln{\frac{\mu^2}{\Lambda^2}}}
+ f\left(\frac{\mu^2}{\Lambda^2}\right) , \;\;\;\; f\left(\frac{\mu^2}{\Lambda^2}\right) =  \frac{4\pi}{\beta_0}\; \frac{ 125\left(1 + 4 \frac{\mu^2}{\Lambda^2}\right)}{\left(1 - \frac{\mu^2}{\Lambda^2}\right)\left(4 + \frac{\mu^2}{\Lambda^2}\right)^4}.
\end{eqnarray}
At low scales it is consistent with global data of power corrections to perturbative observables,  while  for larger values it coincides  with the conventional perturbative running coupling constant with Landau pole as shown in Fig.~\ref{BothRCPIF}.

The final expression used in the numerical analysis is then given by
\begin{eqnarray}
&& \hspace{-1cm}\hat{\alpha}_s \left(Q Q_0, \gamma \right) 
= 
\tilde{\alpha}_s \left(Q Q_0, \gamma \right)
+ \frac{N_c}{\pi} f\left( \frac{QQ_0}{\Lambda^2}\right), \quad 
\end{eqnarray}
which replaces Eq.~(\ref{eq:BLM}) in all expressions. 
In a future publication we will compare the 
scheme here presented to other physical renormalization schemes. For simplicity we have not introduced a complete treatment of quark thresholds in 
the results of this letter, but we have checked that they have a very small 
effect. Let us stress that in our numerical results we do not use any saddle point approximation and perform the numerical integrations exactly.

\newpage

\section{Comparison to DIS data \&  scope}
\begin{figure}[htbp]
  \centering
  \includegraphics[width=\textwidth,angle=0]{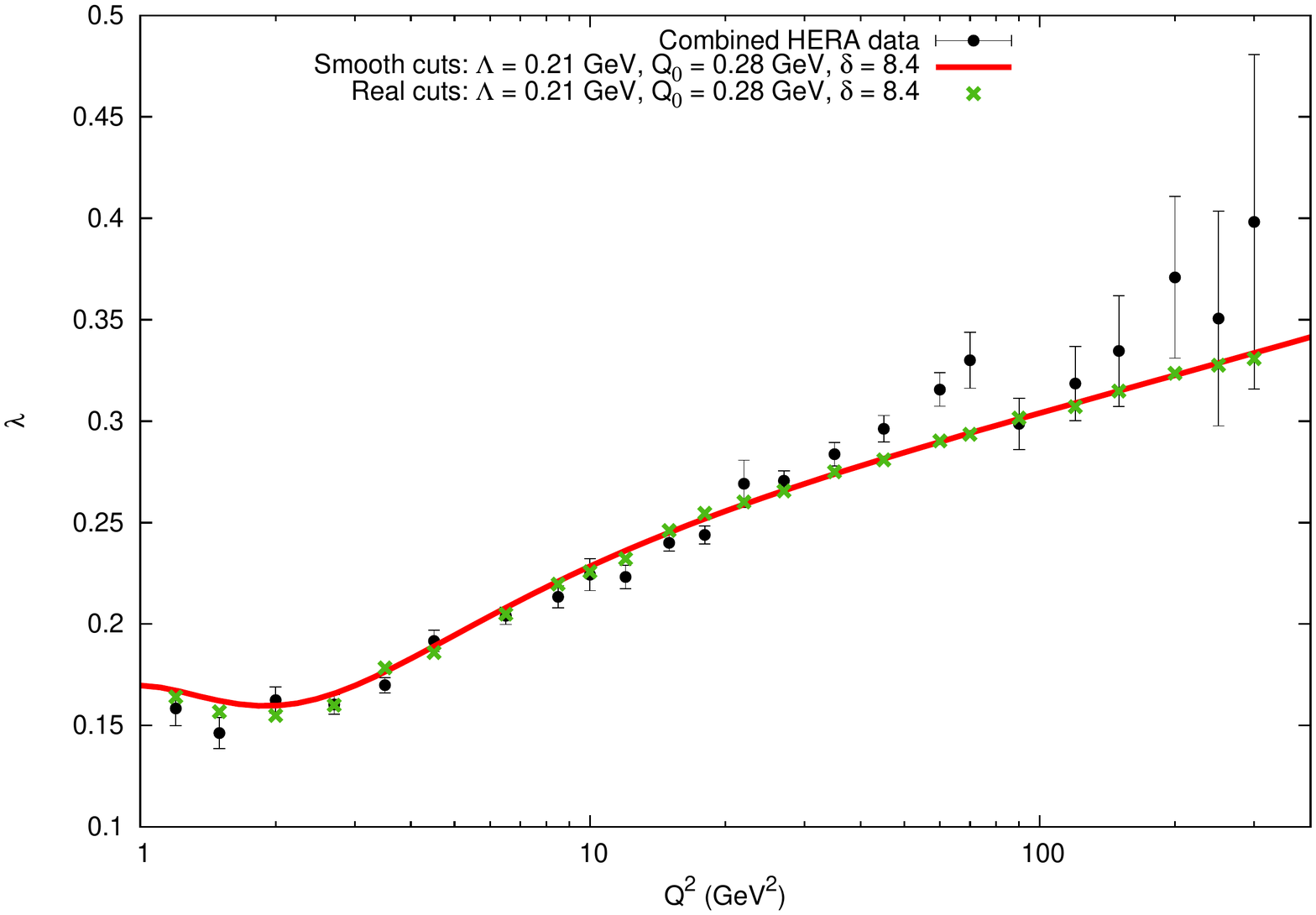}
  \caption{{\small Comparison of our prediction with experimental data.}}
  \label{Hard-Soft-Pomeron}
  \includegraphics[width=.7\textwidth,angle=0]{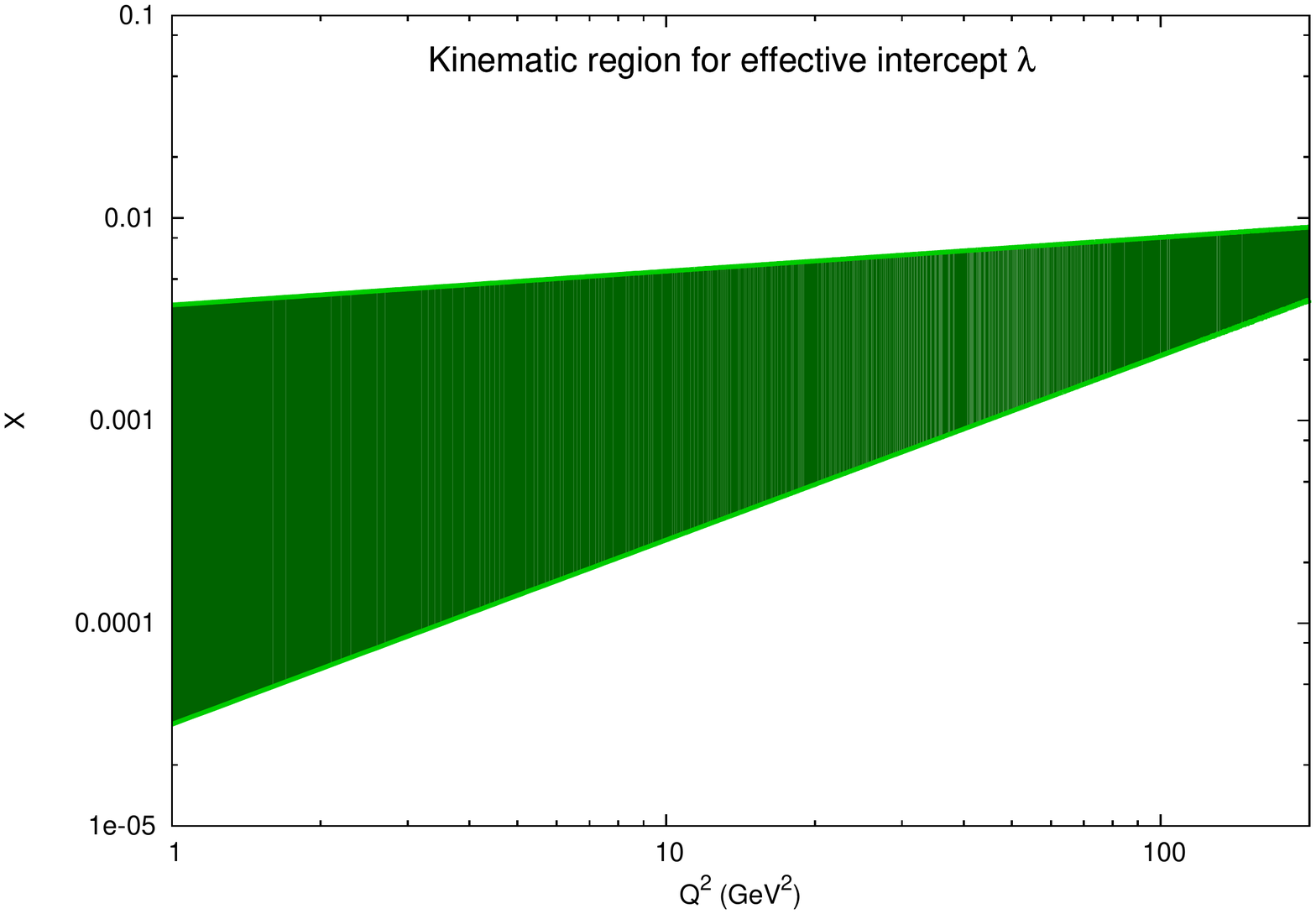}
  \caption{{\small Smooth  cuts in $x$ used for  the effective intercept of $F_2$.}}
  \label{Domain}
\end{figure}

To obtain our theoretical results we have calculated the logarithmic
derivative $\frac{d \log F_2 }{ d \log (1/x)}$ using Eq.~(\ref{Frho}) with the
modifications described in Section~\ref{RunningScheme}.  For the comparison
with DIS data we chose the
values $Q_0 = 0.28 \,{\rm GeV}$ and $\delta = 8.4$ while the dependence on the overall normalization factor  $\mathcal{C}$ cancels for our observable.  The QCD running
coupling constant is evaluated for $n_f=4$ and $\Lambda= 0.21\, {\rm GeV}$,
corresponding to a $\overline{\rm MS}$ coupling of
$\alpha_s^{\overline{\rm MS}}(M_Z^2) = 0.12$. The result is shown in
Fig.~\ref{Hard-Soft-Pomeron}. The experimental input has been derived from the
combined analysis performed by H1 and ZEUS in Ref.~\cite{Aaron:2009aa}
with $x<10^{-2}$. In the results indicated with ``Real cuts" we have
calculated the effective intercept for $F_2$ at a fixed $Q^2$,
averaging its values in a sample of $x$ space consistent with the actual
experimental  cuts in $x$. To generate the continuous line with
label ``Smooth cuts" we have used as boundaries in $x$ space those
shown in Fig.~\ref{Domain}, which correspond to an interpolation of
the real experimental boundaries. Note that the difference between
both approaches is very small.

We would like to stress the accurate description of the combined HERA
data in our approach, in particular at very low values of $Q^2$. It is
noteworthy that the values of $Q_0$ and $\delta$ indicate that our
proton impact factor (see the plot at the right in
Fig.~\ref{BothRCPIF}) safely lies within the non-perturbative region
since it has its maximum at $\sim 0.81 \, {\rm GeV}$.  In the present
letter our intention is to emphasize that, in order to reach the low
$Q^2$ region with a collinearly improved BFKL equation we needed to
call for optimal renormalization and use some model with a frozen
coupling in the infrared. 

It is possible to improve the quality of our
fit by introducing subleading contributions such as threshold effects
in the running of the coupling, heavy quark masses and higher order
corrections to the photon impact factor which became recently
available \cite{Balitsky:2012bs}. We leave these, together with a
comparison to data not averaged over $x$, for a more extensive study,
which will include an investigation of $F_L$, to be presented
elsewhere.

\section*{Acknowledgments} 

We acknowledge partial support from the European Comission under contract LHCPhenoNet (PITN-GA-2010-264564), the Comunidad de Madrid through Proyecto HEPHACOS ESP-1473, and MICINN (FPA2010-17747).  M.H. also acknowledges support from the German Academic Exchange Service
(DAAD), the U.S. Department of Energy under contract number DE-AC02-98CH10886 and a BNL ``Laboratory Directed Research and Development'' grant (LDRD 12-034).

\end{document}